\documentclass[iop]{aastex}
\usepackage{color}
\usepackage{natbib}
\newcommand{\AKARI}{\textit{AKARI}}

\def\Teff{$T_{\rm{eff}}$}
\def\Tcr{$T_{\rm{cr}}$}
\def\Tcond{$T_{\rm{cond}}$}
\def\logg{log \textit{g}}

\def\HtO{$\mathrm{H_{2}O}$}
\def\Ht{$\mathrm{H_{2}}$}
\def\CHf{$\mathrm{CH_4}$}
\def\COt{$\mathrm{CO_2}$}

\def\AKARI{\textit{AKARI}}

\slugcomment{e-mail: sorahana@astron.s.u-tokyo.ac.jp}

\shorttitle{Effect of Elemental Abundances to Near-Infrared Spectra between 1.0 and 5.0~$\mu$m}
\shortauthors{Sorahana et al.}

\begin{document}


\title{{\AKARI} OBSERVATIONS OF BROWN DWARFS. IV. Effect of Elemental Abundances to Near-Infrared Spectra between 1.0 and 5.0~$\mu$m}


\author{S. Sorahana$^{1,2}$ and I. Yamamura$^{2}$}
\affil{$^{1}$Department of Astronomy, Graduate School of Science, The University of Tokyo,
Bunkyo-ku, Tokyo 113-0033, Japan \linebreak
${^2}$Department of Space Astronomy and Astrophysics, Institute of Space and Astronautical Science (ISAS),\\ Japan Aerospace Exploration Agency (JAXA), 
Sagamihara, Kanagawa 252-5210, Japan}
%\email{sorahana@ir.isas.jaxa.jp}


\begin{abstract}
The detection of the {\COt} absorption band at 4.2~$\mu$m in brown dwarf spectra by {\AKARI} has made 
it possible to discuss {\COt} molecular abundance in brown dwarf atmospheres. 
In our previous studies, we found an excess in the 4.2 $\mu$m {\COt} absorption band of three brown dwarf spectra,
and suggested that these deviations were caused by high C and O elemental abundances in their atmospheres. 
To validate this hypothesis we construct a set of models of brown dwarf atmospheres with various elemental abundance patterns, 
and investigate the variations of the molecular composition, thermal structure and their effects to the near-infrared spectra between 1.0 and 5.0~$\mu$m. 

The 4.2~$\mu$m {\COt} absorption band in some late-L and T dwarfs taken by {\AKARI} are stronger or weaker than predictions by corresponding models with solar abundance. 
By comparing {\COt} band in the model spectra to the observed near-infrared spectra, 
we confirm possible elemental abundance variations among brown dwarfs. 
We find that the band strength is especially sensitive to O abundance, but C is also needed to reproduce the entire near-infrared spectra. 
This result indicates that both C and O abundances should increase and decrease simultaneously for brown dwarfs.  
We find that a weaker {\COt} absorption band in a spectrum can also be explained by a model with lower $``$C and O$"$ abundances.  
\end{abstract}





\section{Introduction}
Elemental abundances of low temperature objects such as brown dwarfs and extra-solar giant-gas planets have been studied extensively, for example by \citet{Tsuji_2011}, \citet{Madhusudhan_2011}, \citet{Barman_2011}, \citet{Fortney_2012}, \citet{Konopacky_2013} and \citet{Moses_2013}.
In particular, they focused on the abundances of C and O.
Some researchers including \citet{Madhusudhan_2011}, \citet{Fortney_2012}, and \citet{Moses_2013} especially discussed the C/O ratio in exoplanet atmosphere. The C/O ratio is an important parameter that governs the atmospheric composition, such as abundances of CO, {\HtO}, and {\CHf} molecules. 
The C/O ratio can be an information about the origin and evolution of the object, 
and thus it would be an indicator for the classification of planets.
For example, \citet{Madhusudhan_2011} reported that a hot Jupiter WASP--12b has a carbon rich atmosphere (i.e., C/O $> 1$). 
This indicates a substantial depletion of oxygen in the disk during its evolution.

\citet{Swain_2009} used NASA's Hubble Space Telescope to make the first detection of {\COt} in the spectrum of a transiting extrasolar planet, HD~209458b.
The presence of {\COt} absorption at 4.2~$\mu$m in brown dwarf spectra was first reported by \citet{Yamamura_2010} 
from spectra taken by the Japanese infrared astronomical satellite {\AKARI} (\citealt{Murakami_2007}). 
They found that the {\COt} absorption band at 4.2~$\mu$m in three out of six 
{\AKARI} brown dwarfs was deeper than predicted by the Unified Cloudy Model (UCM; \citealt{Tsuji_2002, Tsuji_2005}), a brown dwarf atmosphere model. 
\citet{Sorahana_2012} investigated the spectral features of 16 {\AKARI} objects and confirmed that the $\mathrm{CO_2}$ molecule is always present in the atmosphere of T dwarfs. 
They also pointed out that the observed $\mathrm{CO_2}$ absorption band in some late-L to T dwarfs is sometimes stronger and sometimes weaker than the predictions by the UCM. 
This problem seems to be related to {\COt} abundance and is not solved by adjusting three model parameters in the UCM; effective temperature {\Teff}, surface gravity {\logg}, and critical temperature {\Tcr} 
(This is an additional parameter in UCM that controls the thickness of the dust cloud in the photosphere: see Section~2 for details). 
\citet{Yamamura_2010} attempted to explain these deviations by vertical mixing following \citet{Griffith_1999} and \citet{Saumon_2000}, 
and concluded that the {\COt} feature could not be reproduced by such mechanism.
Solar elemental abundance (\citealt{Allende_2002}) has been assumed in UCM,  
as it was believed sufficient for analyzing the low resolution spectra of cool dwarfs. 
However, \citet{Tsuji_2011} proposed that the excess of the {\COt} absorption band in the observed spectra found by \citet{Yamamura_2010} could be reproduced by increasing the both C and O abundances. 
In fact, the {\COt} band excess in three brown dwarfs appeared to be reproduced by the C and O abundances of the old solar values (log $A_C$ = 8.60 and log $A_O$ = 8.92; \citealt{Anders_1989, Grevesse_1991}) being larger by about +0.2~dex from the revised solar values (log $A_\mathrm{C}$ = 8.39 and log $A_\mathrm{O}$ = 8.69; \citealt{Allende_2002}).
This result raised other questions; (1) whether abundance variation in brown dwarfs  is a general phenomenon, (2) whether there are objects that have lower C and O abundances, (3) whether only C and O abundances change, and (4) how widely do elemental abundances in brown dwarfs range.

To confirm the validity of the result of \citet{Tsuji_2011} and answer the above questions, 
we construct a set of brown dwarf atmosphere models using the UCM with various elemental abundances different from the standard solar abundance (\citealt{Allende_2002}), 
and investigate the effects on atmospheric structure and infrared spectra from 1.0 to 5.0~$\mu$m.  
We also discuss how are the excess or deficiency in the observed $\mathrm{CO_2}$ absorption band in some late-L and T dwarfs explained.

\section{Brown Dwarf Atmosphere with Various Metallicity}
\label{s2}
In order to test how atmospheric structure and the resultant infrared spectra vary with elemental abundance, 
we take a model of ({\Tcr}/{\logg}/{\Teff})=(1800~K/5.5/1800~K) as an example  L dwarf atmosphere and a model of ({\Tcr}/{\logg}/{\Teff})=(1900~K/4.5/1200~K) as an example T dwarf. 
The physical parameters of these two types of brown dwarfs, {\Teff}, {\logg}, and {\Tcr}, were given by model fitting to the observed spectra of actual objects, SDSS~J0539--0059 (L5) and 2MASS~J0559--1404 (T4.5) \citep{Sorahana_2012}. 

The UCM accounts for dust formation and sublimation/sedimentation.
{\Tcr} is not predictable by any physical theory at present and is an empirical parameter.
We assume that dust (with the size typically $\sim$0.01~$\mu$m) balances with gas and are not growing. 
Then the dust would exist in a layer of {\Tcr} $<T< $ {\Tcond} (see also \citealt{Tsuji_2002, Tsuji_2005} for details).
The UCM applies the line lists of {\COt} (HITEMP database;
\citealt{Rothman_1997}), {\CHf} (\citealt{Freedman_2008} based on the Spherical Top Data System model of \citealt{Wenger_1998}),  
CO (\citealt{Guelachvili_1983, Chackerian_1983}), and {\HtO} \citep{Partridge_1997}.

We adjust the elemental abundances from the solar value by $\pm$0.2~dex. 
The elemental abundances [X/H] of a solar neighborhood star are distributed between --1.0 and +1.0~dex (\citealt{Bodaghee_2003}),  
thus variations of --0.2 and +0.2~dex are reasonably within the common abundance range. 
We investigate cases where the following parameters are varied: 
\begin{itemize}
\item[(1)] all metal abundances 
\item[(2)] only carbon abundance
\item[(3)] only oxygen abundance 
\item[(4)] only Fe abundance 
\item[(5)] C and O abundances
\item[(6)] C and O and Fe abundances.  
\end{itemize}
Table~\ref{allabu} shows the elemental abundances applied in this study. 
For each case we increase ($+0.2$~dex) or decrease ($-0.2$~dex) the abundance, 
therefore we have 12 total test cases. 

``All metal abundances" mean all elemental abundances except for Hydrogen and Helium. 
Carbon and Oxygen are the main components of the major molecules in brown dwarf atmospheres together with Hydrogen. 
Dust components treated in the UCM are Fe, MgSiO$_3$, and Al$_2$O$_3$ \citep{Tsuji_2002}. 
The most important dust component among them is Fe, because it has the highest number density at the low temperatures encountered in brown dwarf atmospheres. 

In the wavelength range of 1.0--5.0~$\mu$m, there are fundamental and overtone bands of major molecules: 
{\HtO} at 1.4, 1.8, and 2.7~$\mu$m; CO at 2.3 and 4.6~$\mu$m; {\COt} at 4.2~$\mu$m; and {\CHf} at 2.2 and 3.3~$\mu$m.  
It is expected that their abundances are affected by those of C and O. 
The $J$, $H$, and $K$ band fluxes are affected mainly by the Fe dust abundance and continuum source intensity. 
In evaluating the calculated models, we focus on the following three points: 
(a) changes in H$_2$O, CO, CO$_2$, and CH$_4$ abundances in the photosphere; 
(b) variation of the temperature structure;   
and (c) molecular band profiles and flux levels in the $J$, $H$, and $K$ bands.   
We show the result of each calculation in the following Sections, \ref{allmetal} -- \ref{cofmetal}. 

\begin{deluxetable}{lrrr}  
\tabletypesize{\scriptsize}
  \tablecaption{Abundances of the 34 most abundant elements used in the current study 
  \label{allabu}}
\tablewidth{0pt}
\tablehead{
\colhead{Elements} & \colhead{Solar$^1$} & \colhead{$+0.2$~dex} & \colhead{$-0.2$~dex} \\  
}
\startdata
H       &  0.00      &   0.00  &   0.00  \\
He     &  --1.01  &        --1.01 & --1.01 \\ 
Li      &--8.69  &       --8.49 & --8.89 \\
Be     & --10.85    &       --10.65 &--11.05 \\   
B          &--9.40 &      --9.20 & --9.60\\
C       &  --3.61 &      --3.41 &--3.81 \\       
N        &--4.00 &         --3.80 &--4.20\\
O        &--3.31 &         --3.11  &--3.51 \\   
F        &  --7.44 &--7.24  & --7.64 \\
Na    &  --5.67 &  --5.47   &--5.87 \\
Mg      &  --4.42 &--4.22  &     --4.62    \\
Al         & --5.53 &--5.33 &    --5.73   \\
Si        & --4.45 & --4.25 &     --4.65      \\
P         &--6.55 & --6.35  &   --6.75    \\ 
S          &--4.79 & --4.59 &     --4.99     \\
Cl       &--6.50 & --6.30 &    --6.70     \\
K         & --6.88  &--6.68    &  --7.08   \\
Ca     &  --5.64 & --5.44  &   --5.84   \\ 
Sc    &  --8.90 &  --8.70 &   --9.10   \\
Ti         &--7.01 &--6.81  &  --7.21 \\
V       & --8.00 & --7.80  & --8.20\\
Cr     &  --6.33 &  --6.13 &  --6.53 \\
Mn       &  --6.61 &  --6.41&    --6.81  \\        
Fe    &   --4.49 & --4.29     &    --4.69     \\   
Ni       &  --5.75 & --5.55  &   --5.95    \\
Cu     &   --7.79 &--7.59  &   --7.99   \\     
Br       &   --9.37 & --9.17  &  --9.57   \\    
Rb      & --9.40 & --9.20 &    --9.60   \\
Sr       & --9.10 &--8.90   &   --9.30     \\   
Y          &--9.76 &--9.56    &   --9.96   \\
Zr    &  --9.40 &  --9.20 & --9.60  \\
I         & --10.49 & --10.29    &   --10.69    \\
Ba       &--9.87 & --9.67    &   --10.07   \\
La      & --10.78  &  --10.58   &    --10.98      
\enddata
\tablenotetext{1}{ Allende Prieto et al. (2002).}
\tablecomments{Values are listed by number relative to hydrogen in logarithmic form for +0.2dex and -0.2dex abundances}
     \end{deluxetable}

\subsection{Models in which All Metal Abundances are Varied}
\label{allmetal}

\begin{figure}
\epsscale{.45} 
\begin{center}
\plotone{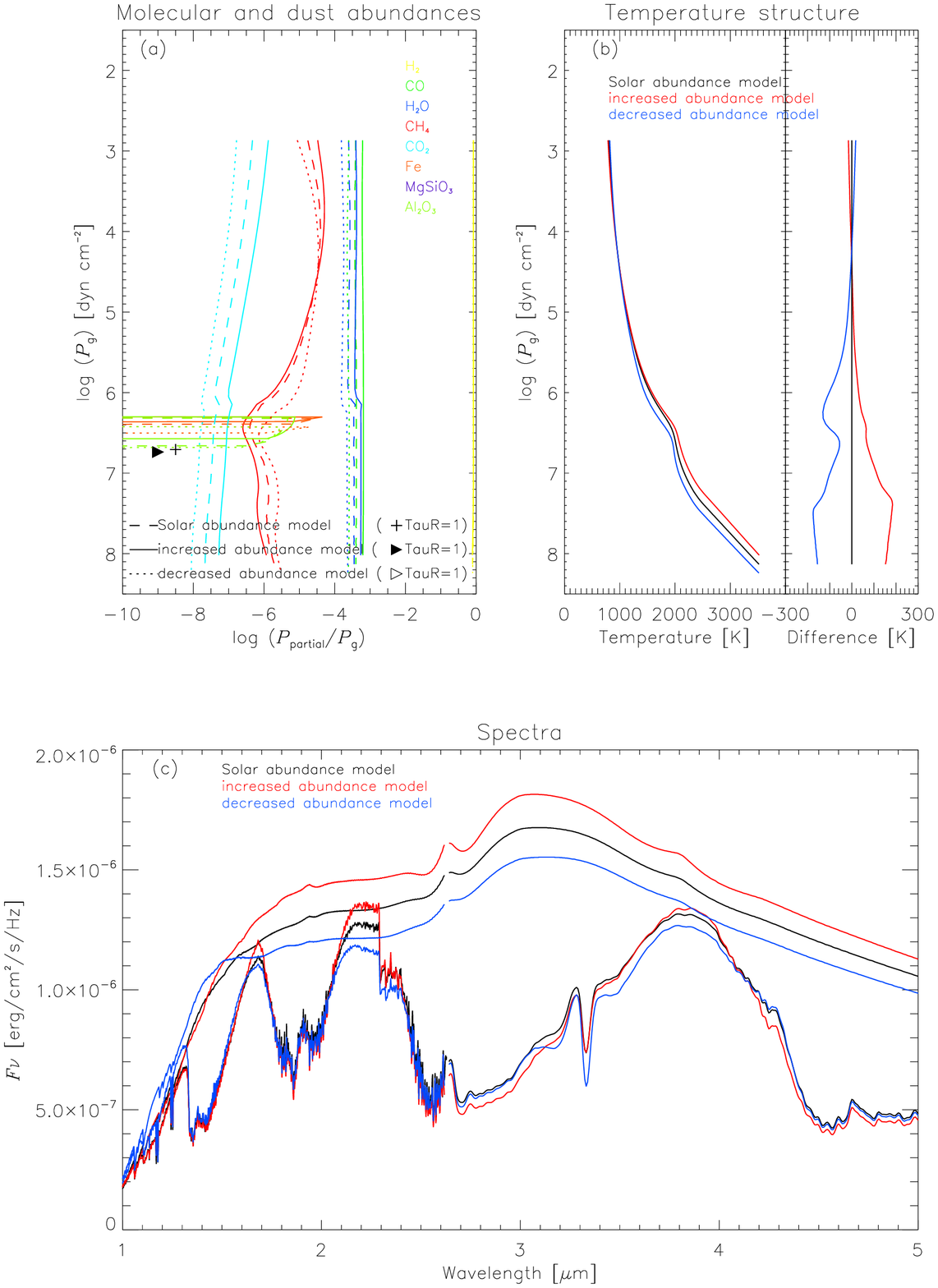}
\end{center}
\caption{Chemical structures, temperature structures, and spectra of the L dwarf model ({\Tcr}/{\logg}/{\Teff}) = (1800K/5.5/1800K) for the cases of increased (solid line for (a) and red lines for (b) and (c)) and decreased (dotted line for (a) and blue lines for (b) and (c)) all metal abundances compared with the solar abundance model. 
(a) Partial pressures of {\Ht} ($\sim$total gas pressure $P_g$), CO, {\HtO}, {\CHf}, and {\COt} molecules normalized by the total gas pressure are plotted against total gas pressure, log $P_g$. 
The abundances of dust are represented by the pressures of the nuclei of the refractive elements locked in the dust grains (i.e., Fe and Al for iron and corundum, respectively).
The position where the Rosseland \& Planck mean opacity becomes unity is also shown; a plus represents the solar abundance model, an open triangle represents the increased abundance case, and a filled triangle represents the decreased abundance case. 
Note that two triangles overlap each other in this figure because the optical depth for the increased and decreased cases are the same. 
(b) \textit{Left:} The temperature structure of each abundance model. 
\textit{Right:} Deviations of the temperature in the models of modified abundance from that of the solar abundance model.
(c) The spectra of the models. The continuum levels of the solar abundance and modified models are also shown. 
} 
\label{La16}
\end{figure}

\begin{figure}
\begin{center}
   \plotone{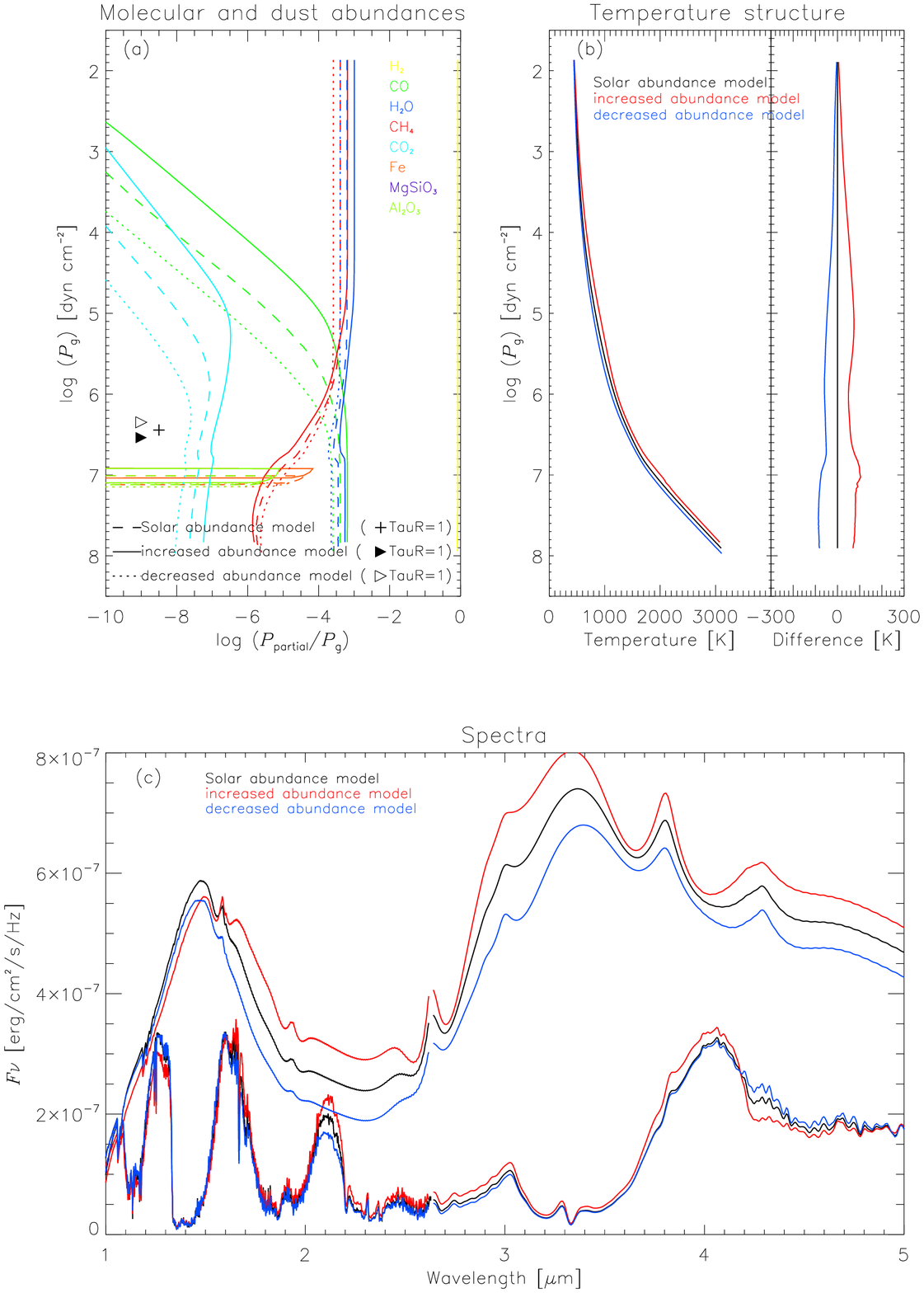}
\end{center}
\caption{T dwarf model ({\Tcr}/{\logg}/{\Teff}) = (1900K/4.5/1200K) with all metal abundances increased. 
The notation is the same as in Figure~\ref{La16}} 
\label{Ta16}
\end{figure}

First, we show the results for the case of increased and decreased abundances of all metal elements (except for H and He) for typical L  (Figures~\ref{La16}) and T dwarfs (Figure~\ref{Ta16}). 
We show the abundance; i.e., partial pressure of each molecule, $\log P_\mathrm{partial}$ [dyn cm$^{-2}$], divided by total gas pressure, $\log P_g$ [dyn cm$^{-2}$], (top left panel); and temperature [K] against total gas pressure, $\log P_g$ [dyn cm$^{-2}$] (top right panel); and the emergent spectrum between 1.0 and 5.0~$\mu$m (bottom panel). 
Partial pressure profiles for the solar abundance (dashed line), increased (solid line), and decreased (dotted line) abundance models are drawn for each molecule in the top left panel. 
We also show the positions at which the Rosseland \& Planck mean opacity becomes unity for each model.
The temperature structures of the modified abundance models (red for increased and blue for decreased models) are also shown as deviations from the temperature of the solar abundance model (black). 
In the bottom panel, the continuum of each abundance model is also shown.

Figure~\ref{La16} shows the L dwarf case. 
For the case of increased elemental abundances, we see that the abundances of all major molecules increase, except for {\CHf} in deep layers ($\log P_g \geq 5$). 
{\CHf} decreases in deep layers by $\sim 0.3$~dex, according to the increased temperature due to increasing {\HtO};
the molecule holds energy from the inner, warmer region by high opacity (greenhouse effect).
On the other hand, surface {\CHf} abundance increases because of enhanced carbon. 
The gas temperature becomes lower toward the surface,  
because of more efficient radiative cooling following the increase of molecular abundances in the optically thin region.
All absorption bands, except for that of {\CHf}, become deeper by 5--7~\% because of the enhanced abundances. 
The flux level around 3.3~$\mu$m {\CHf} band is a result of radiation from the background continuum source traveling through the atmosphere with varying molecular number density and temperature.
Entire flux level along $J$, $H$, and $K$ bands basically reflects the continuum level, which depends on the inner region ($\tau_{R}\sim1$) temperature and dust opacity, as shown in Figure~1 of \citet{Tsuji_2002}.
In this case, dust abundance increases, but inner temperature also increase.
Thus the flux of $J$ band, which is the most affected by dust, does not change, and $H$ and $K$ band fluxes increase. 
For the case of decreased elemental abundances, the change in molecular abundances, temperature structure, and spectral features are generally opposite to the case of increased elemental abundances. 
Almost all molecular abundances decrease, except for the {\CHf} abundance in the inner region ($\log P_g \geq 5$). 
The temperature of the inner region decreases by 150~K because of the {\HtO} abundance decreasing by $\sim 0.2$~dex. 
Consequently, the {\CHf} abundance in this region increases by $\sim 0.3$~dex.  
The surface temperature becomes higher by about 20~K. 
It is caused by the suppressed cooling effect due to the decrease of molecules, opposite to the increased elemental abundance case. Although {\HtO}, {\COt}, and CO abundances decrease by about 0.2--0.5~dex in the optically thin surface area, 
the absorption bands of these molecules barely change.
The $H$ and $K$ band fluxes drop because of the decreasing continuum level due to the lower temperature of the inner optically thick region. On the other hand, $J$ band flux raises because of the raising continuum flux level due to the decreasing dust amount.

In the corresponding T dwarf model shown in Figure~\ref{Ta16}, the abundances of all molecules increase 
almost proportionally to the increased elemental abundances, except for {\CHf} in the inner layers, as observed in the L dwarf model. 
The temperature rises throughout the entire atmosphere. 
The absorption bands of CO and {\COt} become deeper, as observed in the case of L dwarf model. 
The change in the {\COt} band at 4.2~$\mu$m is particularly significant, as much as 20~\%.
On the other hand, the flux levels in the 2.4--3.8~$\mu$m region (including {\HtO} and {\CHf} bands) become higher, contrary to the case of L dwarf model. 
This is because of the rise in the continuum flux level due to the higher photosphere temperature rather than changes of {\HtO} and {\CHf} abundances. 
The changes of the $J$, $H$, and $K$ band fluxes are due to the continuum level.  
For the decreased elemental abundance case, 
the entire temperature drops due to decreasing {\HtO}.  
The effect of radiative cooling in the optically thin region is reduced, as observed in the L dwarf model. 
The changes in the spectral features are opposite to those of the increased elemental abundance case, but not completely.  
For example, the flux level from 3.0 to 4.0~$\mu$m changes only slightly. 
This is the result of a competition between the decreasing continuum flux level due to the lower temperature and 
the weaker absorption bands due to less abundant {\HtO} and {\CHf}. 
The $J$, $H$, and $K$ band fluxes depend mainly on the continuum level.

\subsection{Models in which Only Carbon Abundance is Varied}

\label{cmetalu}
We show the results for the case in which only the C abundance is varied for L (Figures \ref{Lc16}) and T dwarfs (Figure~\ref{Tc16}), respectively. 
In the increased elemental abundance cases, the abundances of CO and {\CHf} molecules increase. 
Oxygen atoms are captured in CO molecules, 
thus the abundance of {\HtO} decreases.  
Molecular abundances of the decreased C abundance case are opposite to the increased C abundance case. 
The residual oxygen atoms, which cannot get carbon atoms, exist as {\HtO}.

The L dwarf model with increased C abundance is shown in Figures \ref{Lc16}. The CO and {\CHf} abundances increase by about 0.2 and 0.9~dex, respectively. 
The {\HtO} abundance deep inside the photosphere (log $P_g > 6$) 
decreases by 0.5~dex and the temperature in this region also becomes lower by more than 140~K. 
The increase of the surface temperature ($\sim 60$ K) correlates with the decrease of {\HtO} ($\sim$1.0~dex) and {\COt} ($\sim$1.0~dex) molecules as the cooling by these molecules becomes less effective. 
We see that the  {\HtO} and {\COt} absorption bands become shallower by 50~\% and 15~\%, respectively. 
On the other hand, {\CHf} band becomes deeper.
The flux level around the CO band seems mostly unchanged because of balances between changes by CO abundance, continuum level, and surface temperature.
Since the inner region temperature goes down,  
the entire continuum level become lower and the $J$, $H$, and $K$ band fluxes reduce by about 20~\%.
The trends of the decreased C abundance case are generally opposite to those of the increased C abundance case, 
even though the variations in molecular abundances, photosphere temperature, and spectral features are less significant than those of the increased C case. 
Both CO and {\CHf} decrease by about 0.2 and 0.4~dex, respectively, while the {\HtO} abundance increases by about 0.2\,dex.  
Although {\COt} abundance changes very little,
the flux level around 4.2~$\mu$m near the {\COt} absorption band becomes fainter because of the effect of increasing {\HtO} abundance.
The decrease of the surface temperature is caused by the larger cooling effect due to the increased {\HtO} abundance.  
The temperature deep inside increases by about 50~K due to the increase in  
{\HtO} molecules (greenhouse effect).
Because of this increased inner region temperature, 
the continuum flux rises and the fluxes in the $J$ and $H$ bands increase by $\sim 5--10$~\%. 
A slight increase of the continuum level and abundant {\HtO} changes the $K$ band flux a little.

In the T dwarf model shown in Figure~\ref{Tc16}, 
the change of the molecular abundance is not similar to the L dwarf model, especially for the {\HtO} and {\COt}. 
For both the increased and decreased abundance cases, the change of {\HtO} abundance is less significant than those of the L dwarf model especially in the surface region. 
As noted at the beginning of this subsection, the variation of {\HtO} abundance relates to that of CO.
Since the carbon atoms are transferred from CO to {\CHf} as spectral type changes from L to T and temperature decreases, 
the abundance of CO in the T dwarf photospheres is generally smaller than that in the L dwarf photospheres following the result of chemical equilibrium calculation.
Thus the variation of CO in the varied C abundance model is small, and then the change of {\HtO} is also less in T dwarf atmosphere. 
Consequently, the flux levels of the {\HtO} bands at 1.4, 1.8, and 2.7~$\mu$m do not change as much as in the L dwarf model. 
{\COt} abundance increases opposite to the L dwarf model. 
This indicates that {\COt} can be created easer in T dwarf atmosphere than that in L dwarf atmosphere.
However, because of little change of {\COt} abundance and little absolute amount, the {\COt} band also does not change for both the increased and decreased C abundance cases.
The {\CHf} absorption band around 3.3~$\mu$m becomes deeper by $\sim$5~\% corresponding to the increasing {\CHf} abundance. 
The flux level around CO band also becomes lower not only by increasing CO, but also by lower continuum level.
For the decreased abundance case, the flux level of 3.0--4.0~$\mu$m region becomes higher, 
in contrast to the flux level of the same wavelength range in the L dwarf model spectrum being lower than the solar abundance model. 
This is explained as follows. 
In the L dwarf photosphere, the increases of the {\HtO} abundance lowers the flux level. 
In T dwarfs, {\CHf} abundance decreases.  
This effect weakens the absorption band at 3.0 -- 4.0~$\mu$m by 5--20~\%.
The $J$, $H$, and $K$ band fluxes do not change for both the increased and decreased abundance models, 
because the {\HtO} abundance near the surface does not change, and there is little change in the continuum level as the dust effects are negligible in T dwarf atmospheres.

\subsection{Models in which Only Oxygen Abundance is Varied}

\label{ometalu}
\begin{figure}
\begin{center}
\plotone{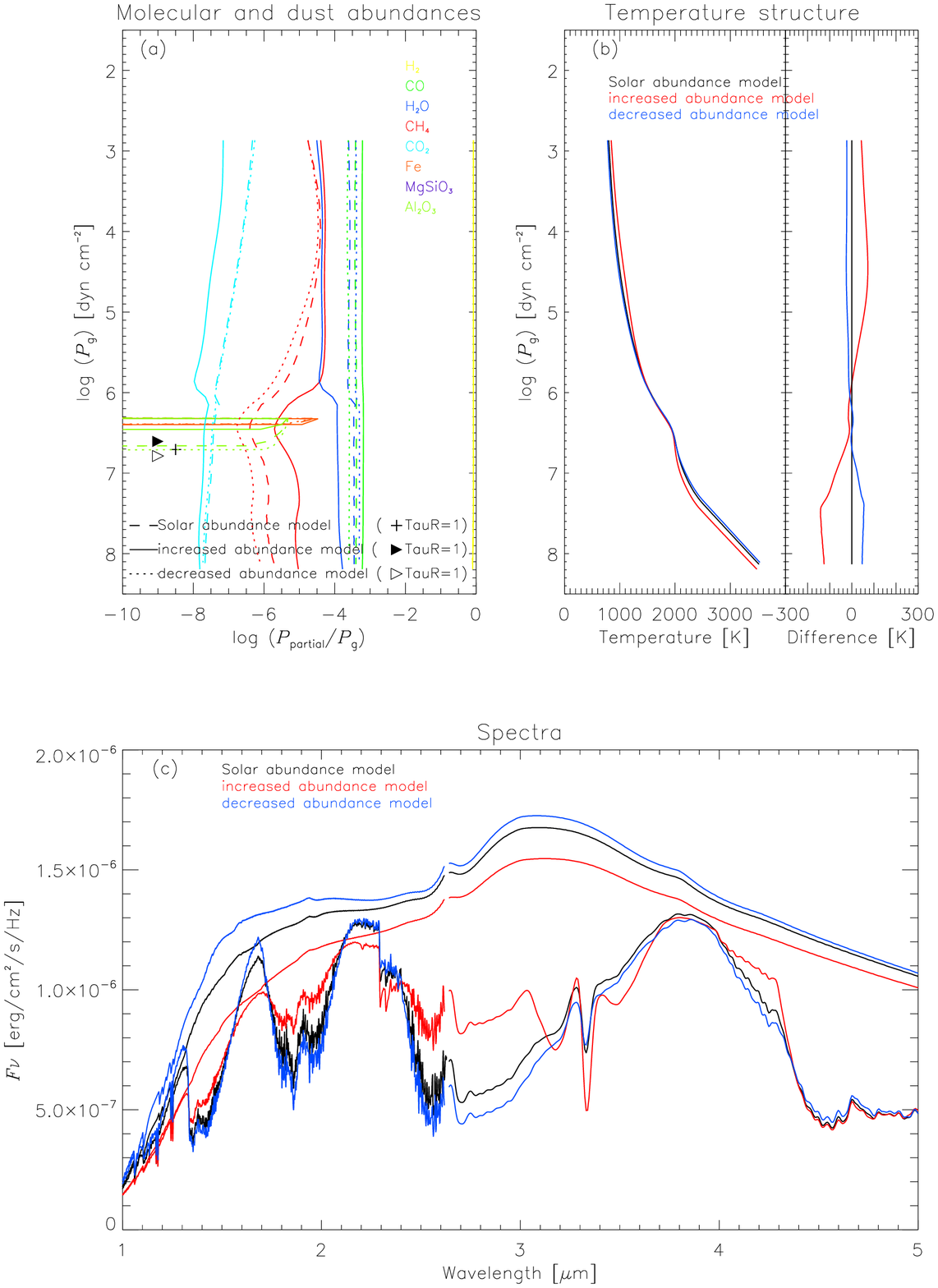}
\end{center}
\caption{
L dwarf model with only C abundance increased/decreased.
The notation is the same as in Figure~\ref{La16}} 
\label{Lc16}
\end{figure}

\begin{figure}
\begin{center}
\plotone{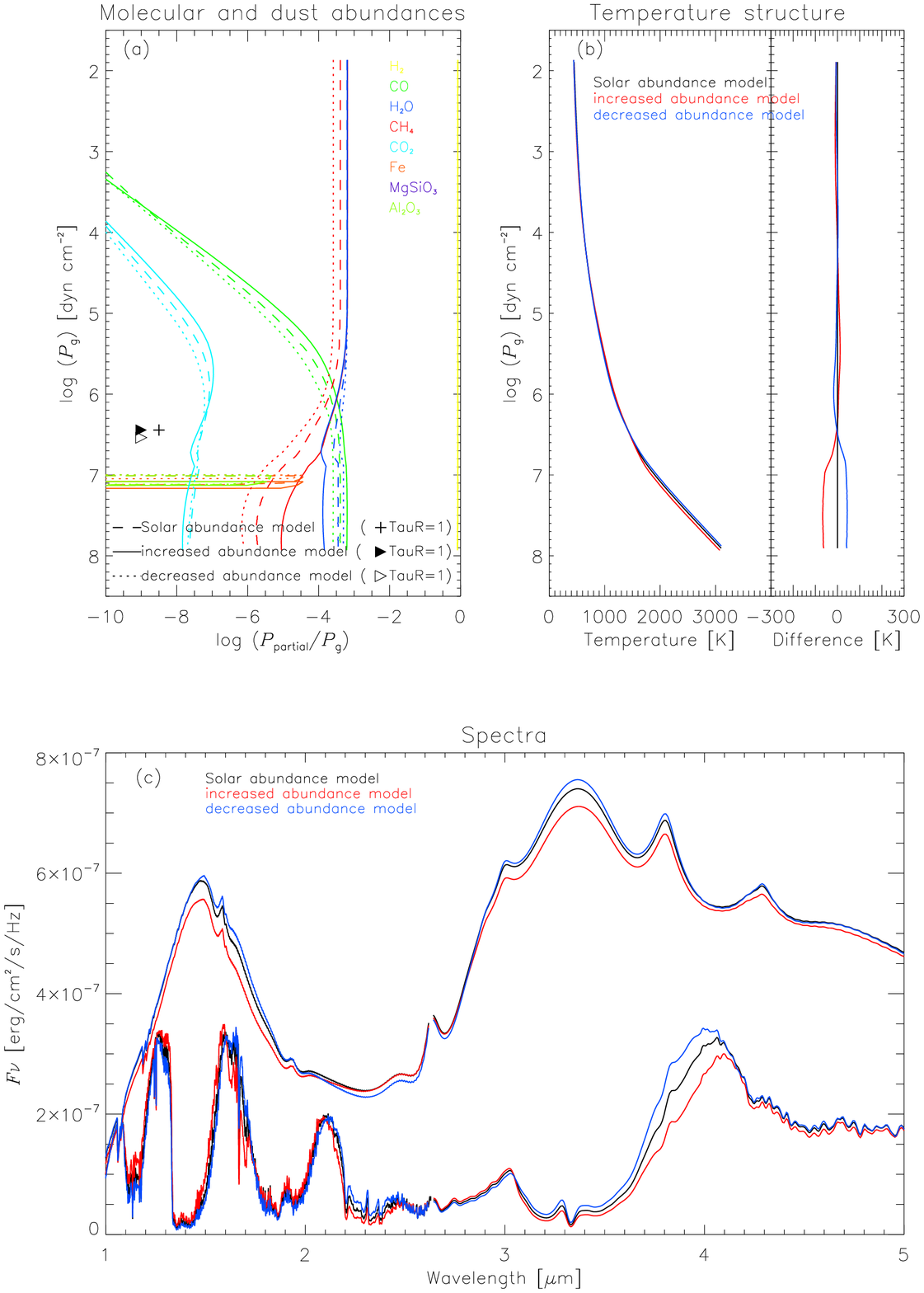}
\end{center}
\caption{ 
T dwarf model with only C abundance increased/decreased.
The notation is the same as in Figure~\ref{La16}} 
\label{Tc16}
\end{figure}

\begin{figure}
\begin{center}
   \plotone{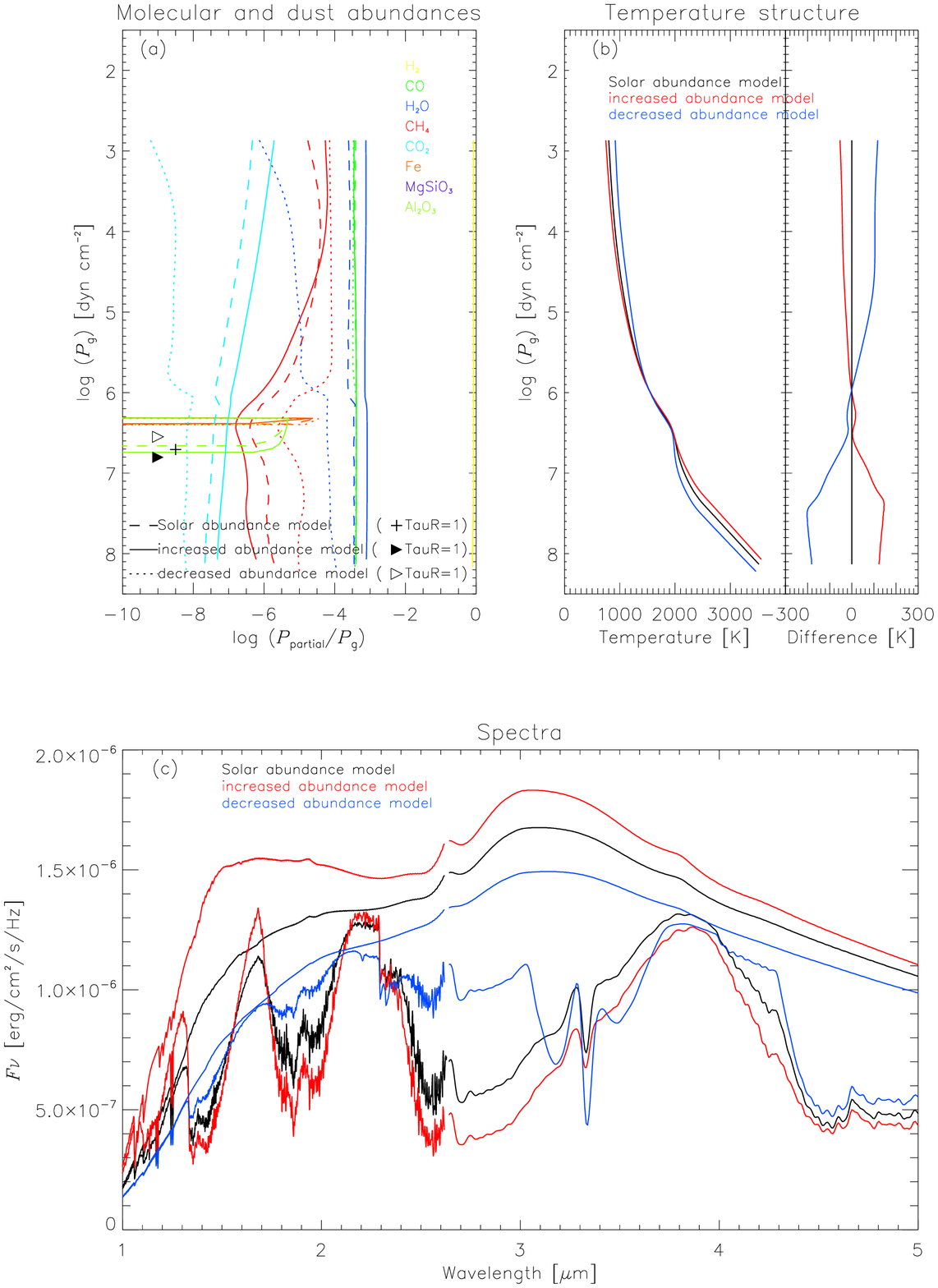}
\end{center}
\caption{
L dwarf model with only O abundance increased/decreased.
The notation is the same as in Figure~\ref{La16}} 
\label{Lo16_2}
\end{figure}

\begin{figure}
\begin{center}
   \plotone{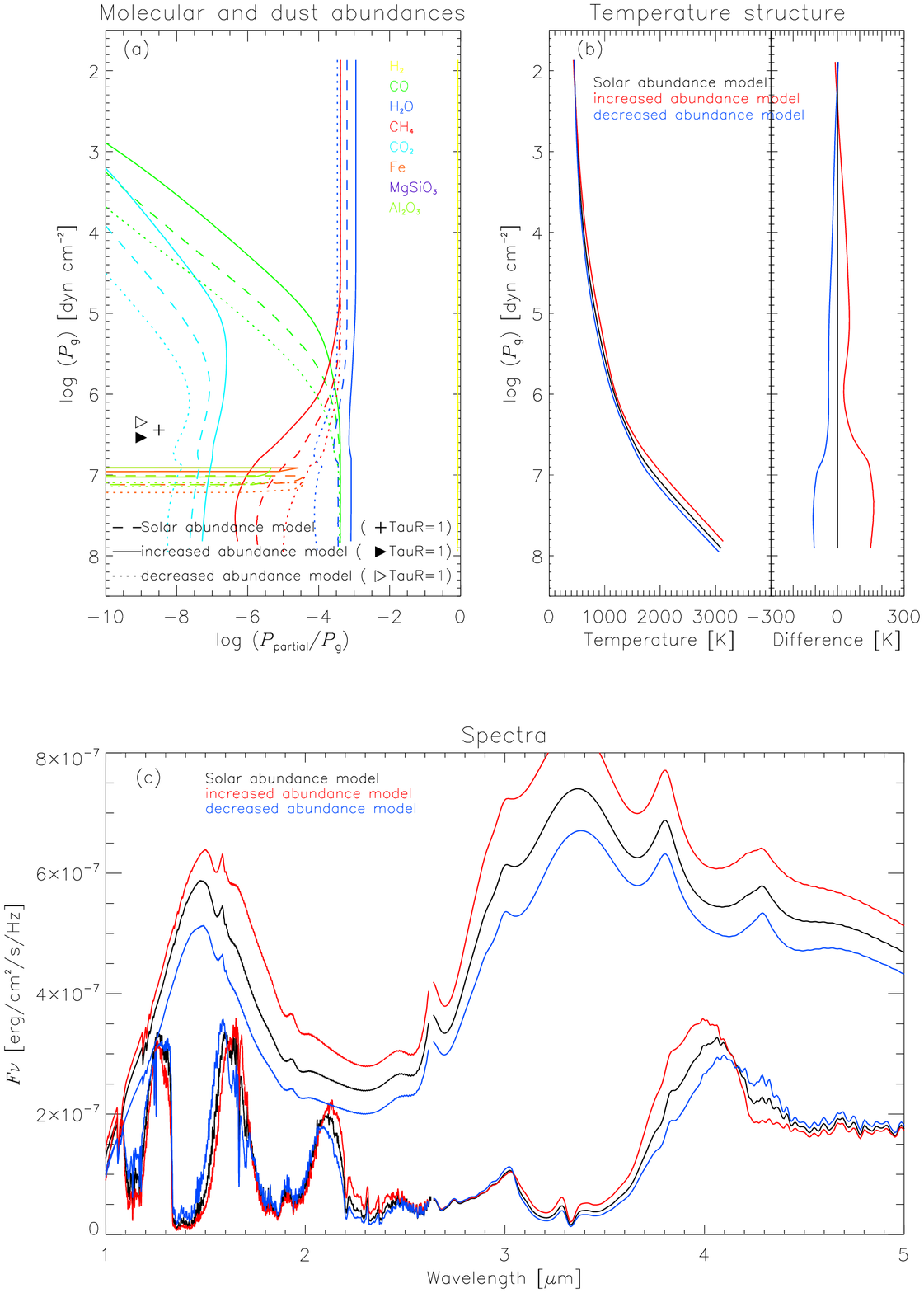}
\end{center}
\caption{
T dwarf model with only O abundance increased/decreased.
The notation is the same as in Figure~\ref{La16}} 
\label{To16}
\end{figure}

The results of the case in which only O abundance is varied in L and T dwarfs are shown in Figures \ref{Lo16_2} and \ref{To16}, respectively. 
In the L dwarf model, the change of {\HtO} abundance and temperature structure in increased/decreased only O abundance model are in principle similar to the case of decreased/increased only C abundance.
Thus the {\HtO} absorption band shape of increased/decreased only O abundance case looks similar with that of decreased/increased only C abundance case. 
In the increased abundance model, 
the {\HtO} and {\COt} abundances increase by about 0.5~dex and the temperature of the inner region ($\log P_g \geq 6.0$) increases by 150~K. 
These variations are larger than those of the decreasing C abundance only model (Section~\ref{cmetalu}; Figure~\ref{Lc16}), 
reflecting the fact that the absolute number of extra O atoms is about double in the case of increasing O abundance.
At the surface region ($\log P_g \leq 5.0$) the temperature decreases by about 50~K due to enhanced cooling by abundant molecules such as {\HtO} and {\COt}. 
The abundance of {\CHf} reflects the temperature structure.
The effects of the increased {\HtO} and {\COt} abundances appear in the model spectra as the deeper absorption bands. 
Although CO abundance change slightly, the flux level around 4.6~$\mu$m CO band becomes lower because of decreased surface temperature.
Because of the rising inner region temperature and decreasing iron grains, the fluxes in the $J$, $H$, and $K$ bands increase by 2--30~\%. 
For the decreased O abundance model, the results are generally in the opposite direction to the case of increased O abundance. {\HtO} and {\COt} are significantly reduced, and the spectral features of such molecular bands dramatically change.

For T dwarf case  shown in Figure~\ref{To16}, the change of {\HtO} and {\CHf} at the surface region in increased/decreased O abundance model is different from that of decreased/increased C abundance model, in contrast to the L dwarf case.
In the only C abundance varied model, carbon abundance contributes to CO and {\CHf} abundances, and then CO abundance affects to {\HtO} abundance. 
On the other hand, in the only O abundance varied model, oxygen abundance directly affect to {\HtO} abundance,
and {\CHf} abundance reflects temperature structure that depends on the {\HtO} abundance.
The flux level around 3.3~$\mu$m {\CHf} absorption band in both the increased/decreased only O abundance is similar to the decreased/increased only C abundance model.
Since {\CHf} abundance at surface region does not change in the model of varied O abundance only, 
we recognize that the 3.3~$\mu$m {\CHf} absorption band reflects the abundance of {\CHf} not only in the surface region but also inner region.
The variations of CO and {\COt} abundances of the only O abundance varied case are different from those of the varied only C abundance model, even though the directions of the changes are the same to each other.
This indicates that both CO and {\COt} abundances directly reflect the abundances of both carbon and oxygen, but more strongly to the O abundance.
These variations appear in spectrum of varied O elemental abundance model.
Especially, the spectral feature of {\COt} band significantly changes.
The temperature increases and decreases in the almost entire region for increased and decreased abundance case, respectively. 
Because of this, the flux in 2.5--3.0~$\mu$m stays at the same level even though {\HtO} abundance  increases/decreases, unlike the case of the L dwarf model.

\subsection{Models in which Only Fe Abundance is Varied}
\label{fmetal}

\begin{figure}
\begin{center}
   \plotone{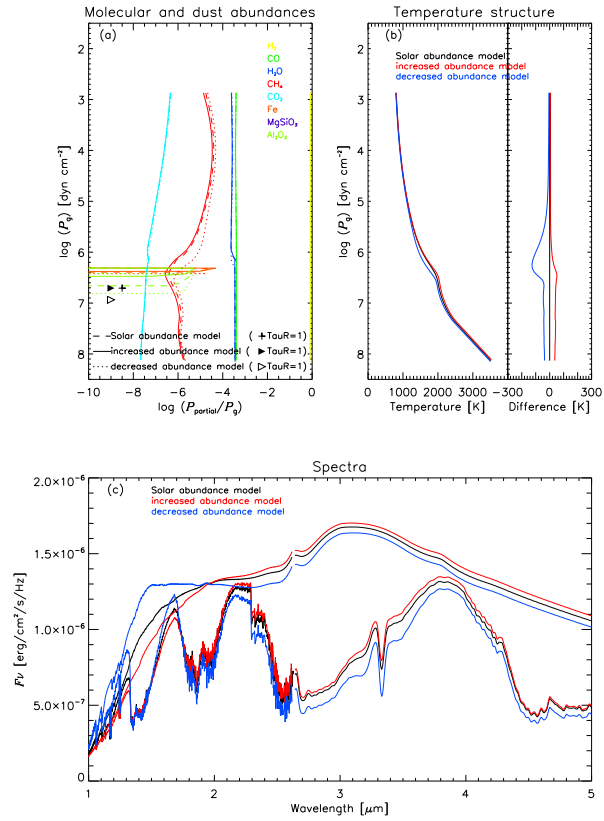}
\end{center}
\caption{ 
L dwarf model with only Fe abundance increased/decreased.
The notation is the same as in Figure~\ref{La16}} 
\label{Lf16_2}
\end{figure}

\begin{figure}
\begin{center}
   \plotone{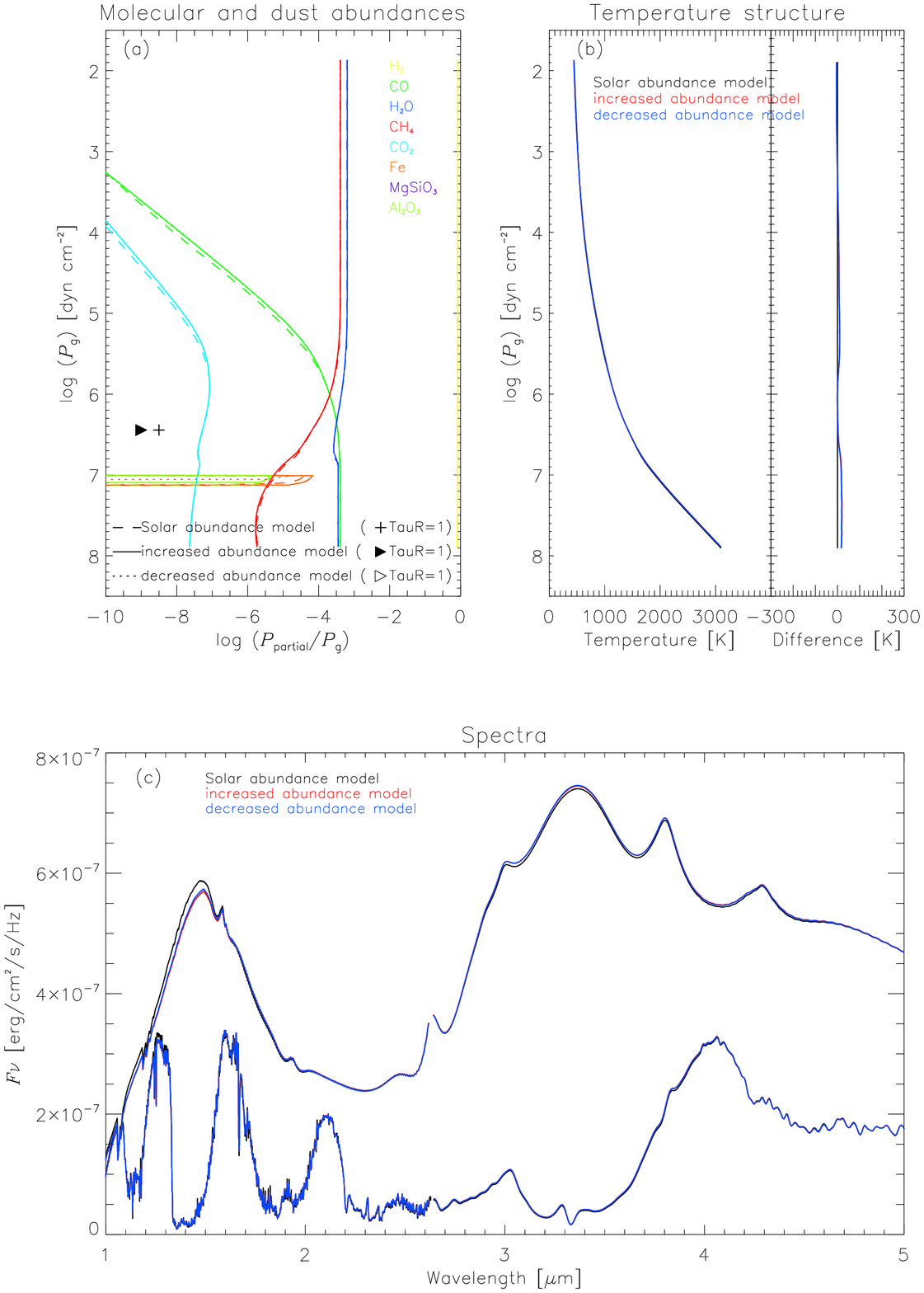}
\end{center}
\caption{
The T dwarf model with only Fe abundance increased/decreased.
The notation is the same as in Figure~\ref{La16}} 
\label{Tf16}
\end{figure}

The results of varied Fe abundance only for L and T dwarfs are shown in Figures \ref{Lf16_2} and \ref{Tf16}, respectively. 
In the L dwarf model with increased Fe abundance, the temperature inside the dust layer ($\log P_g \sim 6.5$) increases by about 50~K, in line with the increase of the amount of Fe dust by 0.5~dex. 
Subsequently, the abundance of {\CHf} decreases slightly in the dust layers. 
The fluxes in the $J$ and $H$ bands decrease 5--13~\% due to larger dust extinction, 
while the flux level in 2.0--4.0~$\mu$m  
increases by about 5~\% reflecting the rising continuum level and lower {\CHf} abundance. 
The case of decreased Fe shows a almost opposite trend to the case of increased Fe. 

On the other hand, the partial pressure of each molecule, and the temperature of the T dwarf model shown in Figure~\ref{Tf16} change only little, 
and the spectrum stays identical to the solar abundance model. 
This is expected, as dust in a T dwarf photosphere precipitates deep inside 
and does not affect the spectrum.

\subsection{Models in which C and O Abundances are Varied}
\label{cometal}

\begin{figure}
\begin{center}
   \plotone{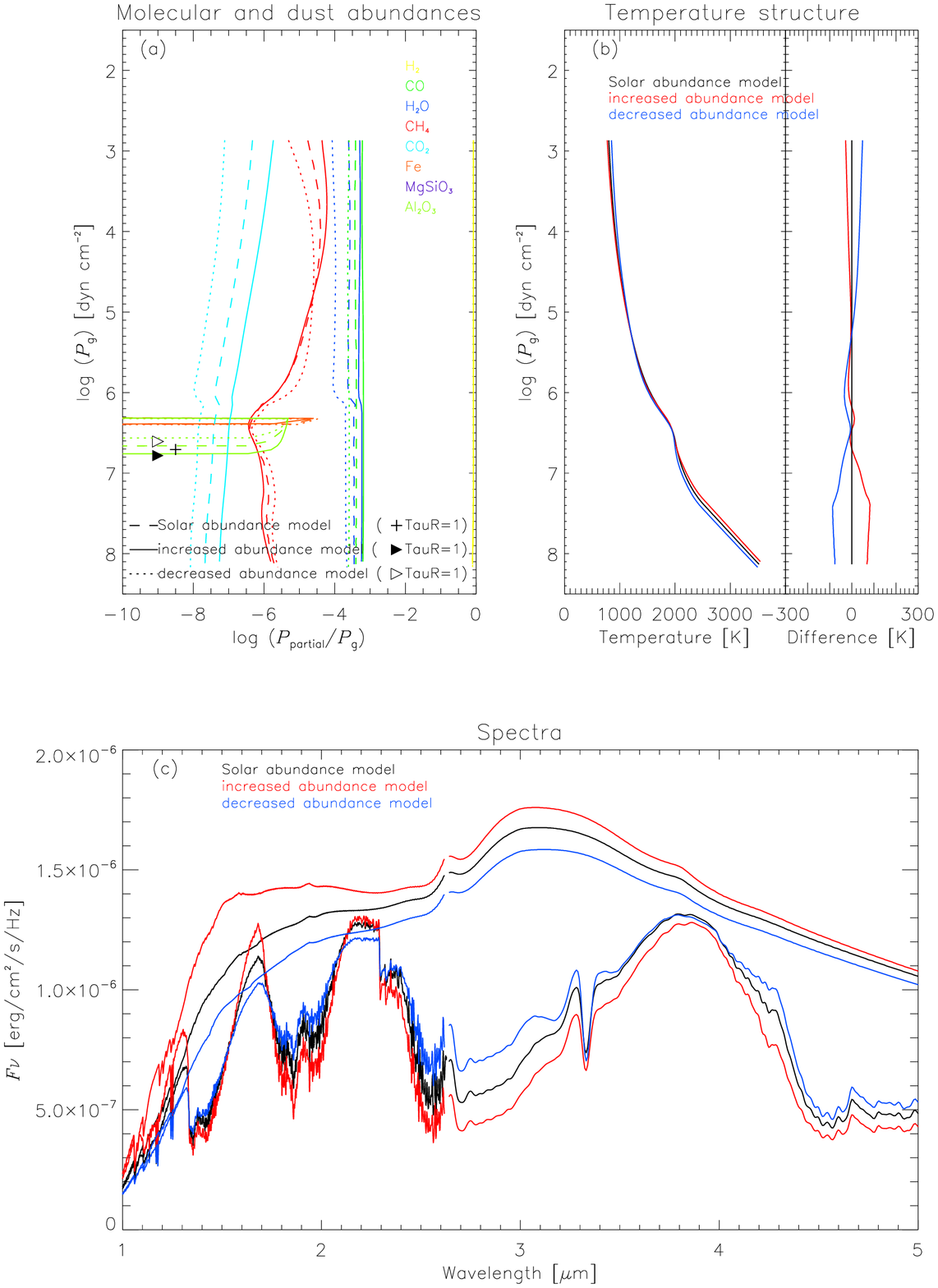}
\end{center}
\caption{ 
The L dwarf model with C and O abundances increased/decreased.
The notation is the same as in Figure~\ref{La16}} 
\label{Lco16}
\end{figure}

The result of varied C and O abundances of the L dwarf model is shown in Figure \ref{Lco16}. 
In the increased C and O abundance model,
the spectral features are generally the same as that of the model with only increased O abundance (Section~\ref{ometalu}; Figure~\ref{Lo16_2}), even though the variation of spectral features is generally smaller than that of the model with only increased O abundance. 
We can understand this as competition between the increasing C and increasing O. 
The abundances of {\HtO}, CO, and {\COt} increase by 0.4, 0.3, and 0.7~dex, respectively. 
These changes make the temperature in the inner photosphere higher by about 80~K, 
and the surface temperature lower by 30~K. 
A gradual decrease of temperature towards the surface ($\log P_g \sim 3$) is caused by the same reason as the case in which only O abundance is increased, 
i.e., the energy from the inner region is transferred to outside more efficiently by the more abundant molecules.  
Following the change of the thermal structure, the abundance of {\CHf} in the inner region of the photosphere decreases by 0.2~dex, but increases by 0.5~dex near the surface. 
The flux level between 2.3 and 5.0~$\mu$m is diminished by 2--30~\% because of increased {\HtO}, CO, {\COt} and surface {\CHf} abundances.  
The {\COt} abundance increases significantly throughout the entire photosphere, and the {\COt} absorption band becomes deeper by about 20~\%. 
The flux in the $J$ and $H$ bands increase by 15--20~\% because of  the rising continuum flux level.  
The $K$ band flux also slightly increases due to the higher temperature in the inner photosphere. 
The decreased abundance case is in general opposite to the increased case. 
The variation of spectral features is generally smaller than that of the model with only decreased O. 
The abundances of {\HtO}, CO, and {\COt} decrease by 0.5, 0.2, and 0.8~dex, respectively. 
The temperature of the inner photosphere becomes lower by 80~K and the {\CHf} abundance increases by 0.2~dex. 
The higher surface temperature is caused by the decreased {\HtO} and {\COt} as in the case of decreased only O abundance. 
Due to the lower inner region temperature, the flux values in the $J$, $H$, and $K$ bands decrease by 5--15~\%. 

\begin{figure}
\begin{center}
  \plotone{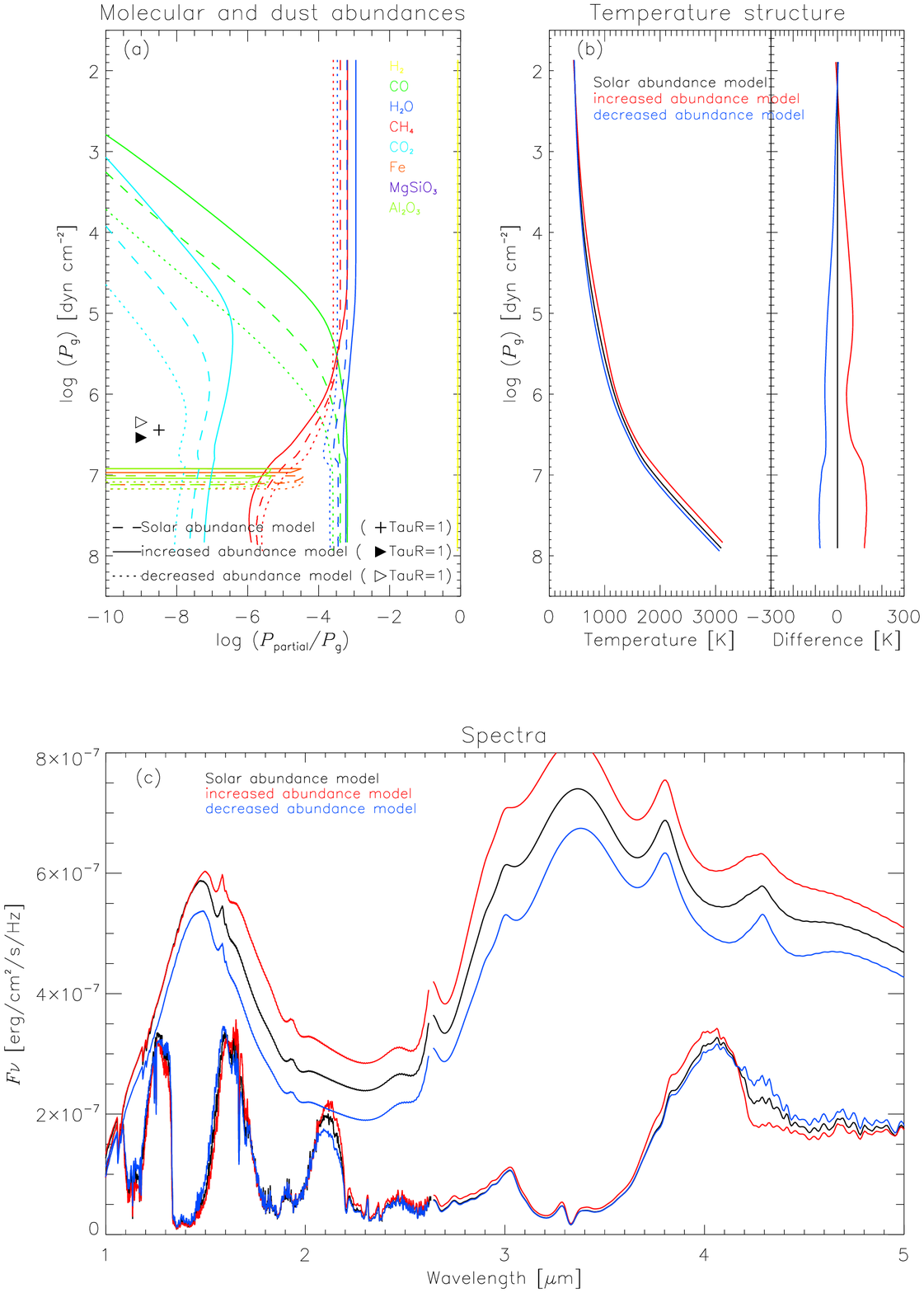}
\end{center}
\caption{
The T dwarf model with C and O abundances increased/decreased.
The notation is the same as in Figure~\ref{La16}} 
\label{Tco16}
\end{figure}

The T dwarf models are shown in Figure~\ref{Tco16}. 
In general, the trend of the model atmosphere for the increased C and O abundances case is similar to that of the increased only O abundance case (Section~\ref{ometalu}; Figure~\ref{To16}), except for the surface {\CHf} abundance.  
However, the variation of the spectral features is smaller to the L dwarf case. 
Since {\CHf} increases at the surface in the model of varied C and O abundances, flux level around {\CHf} absorption band becomes lower than that in the spectrum of increased only O abundance model.
The change in the {\COt} band is significant compared to the variations of other features.   
This indicates that changes of the C and O abundances in the T dwarf model mostly appear in the {\COt} absorption band in the spectrum. 
These results confirm the suggestion by \citet{Tsuji_2011} that {\COt} can be an index of C and O abundances of brown dwarfs, especially for T dwarfs. 
The temperature of the photosphere of the decreased C and O abundance model drops by about 80 K, but the spectrum does not change much, 
except for the shallower 4.2~$\mu$m {\COt}, 4.6~$\mu$m CO bands and smaller $K$ band fluxes. 
The {\COt} and CO band strengths are affected by the changes in their abundances. 
The $K$ band flux level reflects the decreasing continuum flux due to the decreasing inner temperature. 
The decreased continuum level is compensated by the shallower {\HtO} and {\CHf} absorption bands, 
resulting in minor changes to the 2.5--4.0~$\mu$m spectrum.

\subsection{Models in which C, O and Fe Abundances are Varied}
\label{cofmetal}

\begin{figure}
\begin{center}
   \plotone{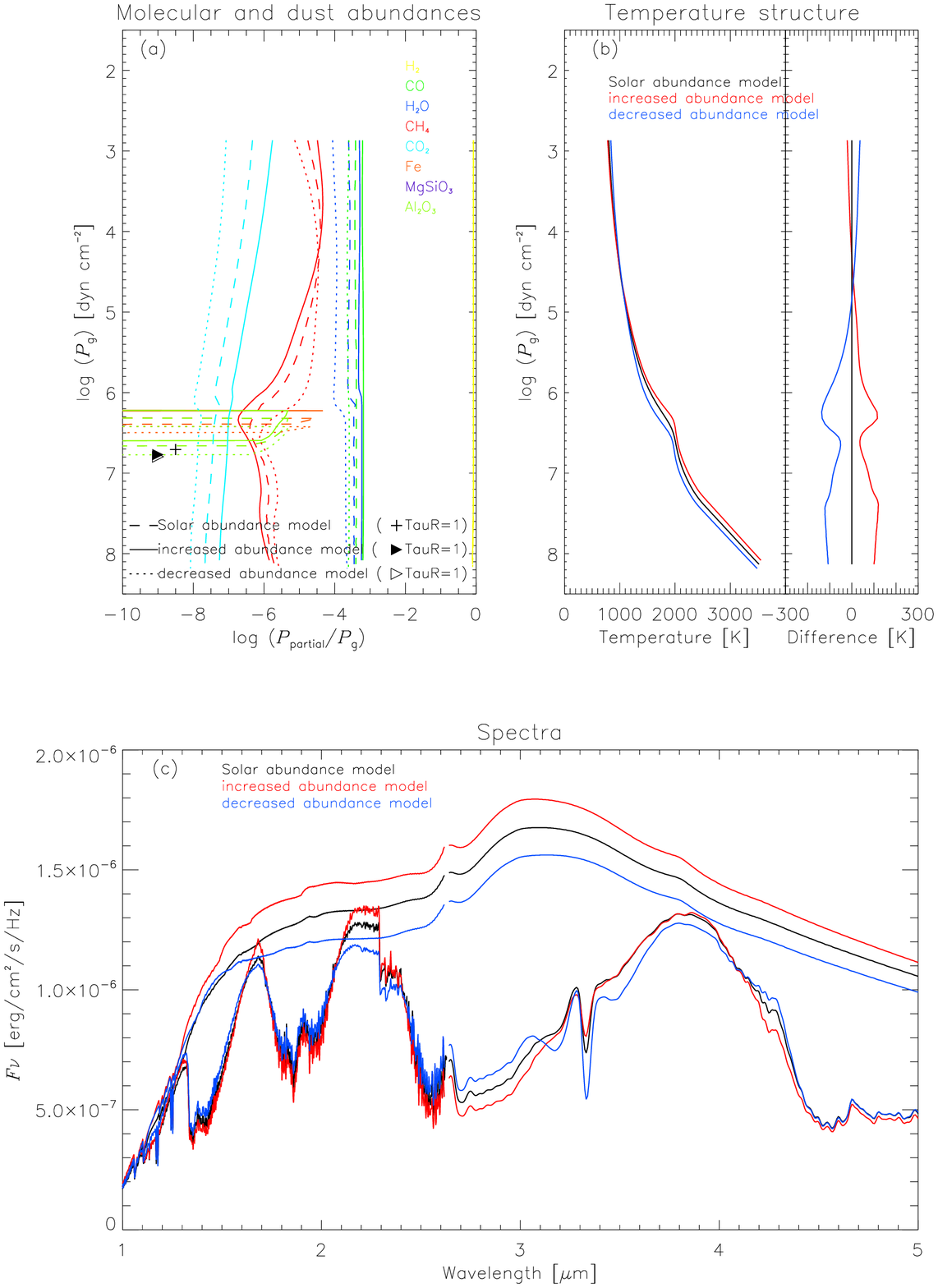}
\end{center}
\caption{ 
The L dwarf model with C, O and Fe abundances increased/decreased.
The notation is the same as in Figure~\ref{La16}} 
\label{Lcof16_2}
\end{figure}

\begin{figure}
\begin{center}
   \plotone{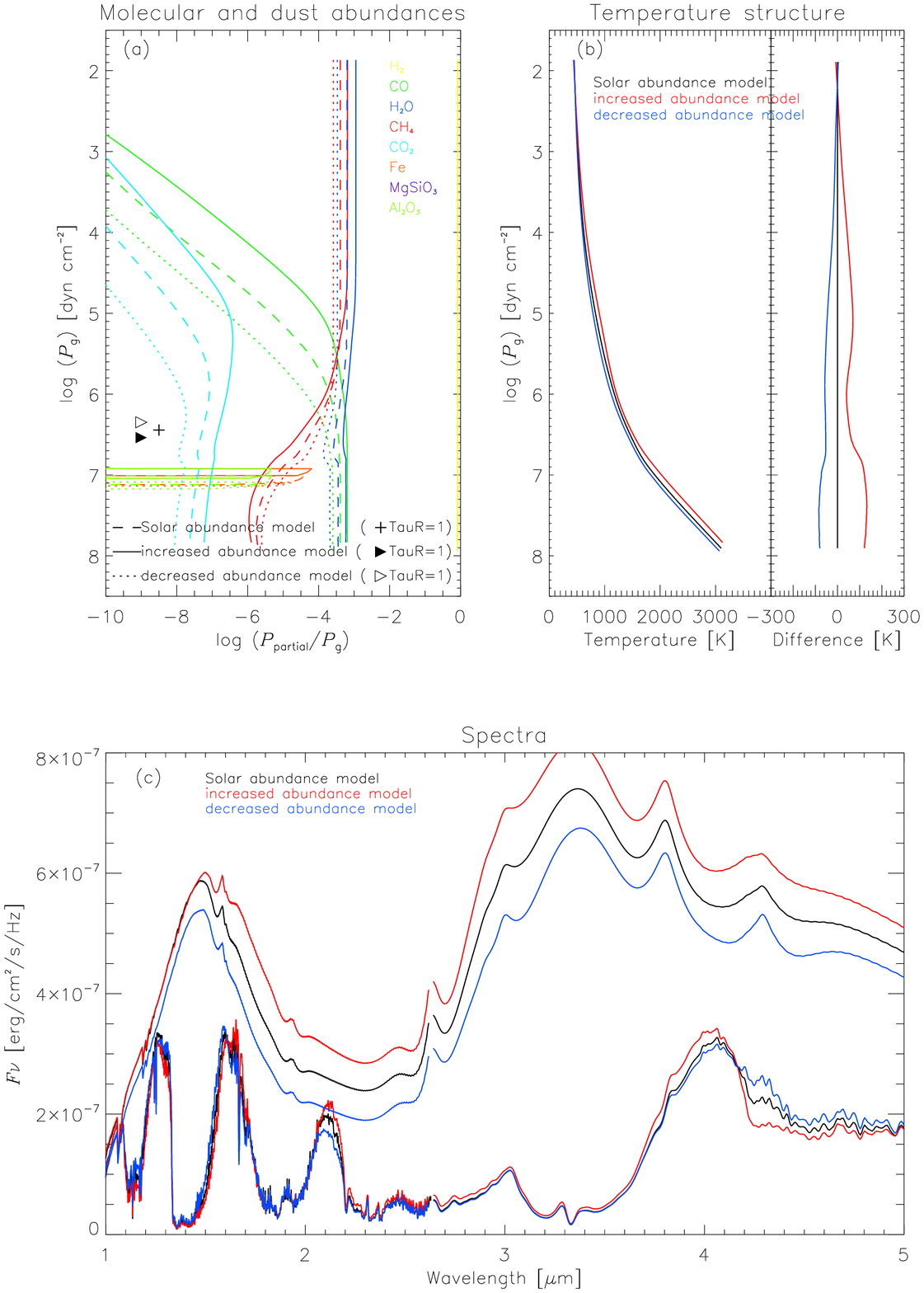}
\end{center}
\caption{ 
The T dwarf model with C, O and Fe abundances increased/decreased.
The notation is the same as in Figure~\ref{La16}} 
\label{Tcof16}
\end{figure}

The results of cases in which C, O, and Fe abundances are varied in L and T dwarfs are shown in Figures \ref{Lcof16_2} and \ref{Tcof16}, respectively. 
In the cases of both increased and decreased C, O, and Fe in the L dwarf model,
the change of the abundances of {\HtO}, CO, and {\COt} are almost the same with those of the models in which C and O abundance are varied.
On the other hand, {\CHf} abundance, temperature structure, and spectral features are different from those of the models in which C and O abundance are varied.
For increased case, more abundant Fe enhances dust formation, and the temperature, especially around the dust layer ($\log P_g \sim 6.2$), increases by more than 100~K. 
The higher temperature raises the continuous flux level and smaller {\CHf} abundance.
Higher continuum level compensate the deepened {\HtO} absorption. 
Consequently, the variation of entire flux level seems to be smaller than that of the case in which C and O abundance are varied.
The fluxes in the $J$ and $H$ bands are affected by the dust amount along with the effects of the inner temperature as discussed in the model spectrum of varied C and O abundances.
In the decreased abundance case, the variations in molecular abundance and temperature structure are also similar to the case of decreased C and O abundances, 
but the decrease of temperature around the dust layer in the current model is larger than that in the case of decreased C and O abundances as is the case of increased abundance. 
Thus the variation of continuum flux level in the current model is also large.
Accordingly, the decrease of flux level over the entire wavelength range is moderate compared to the case of decreased C and O abundances, except for {\CHf} absorption band. 
Since {\CHf} abundance increases above the dust layer due to the lower temperature, the depth of this absorption band becomes deeper.
The $K$ band flux reflects the lower continuum level due to the decreased inner region temperature. 

The changes in the molecular abundances except for Fe, temperature, and spectrum of the T dwarf model (Figure~\ref{Tcof16}) with increased C and O and Fe abundances are almost identical with the case of increased C and O abundances. 
This can be understood from the fact that  dust has little effect on the T dwarf model.
The change in the {\COt} absorption band is particularly large as observed in the varied C and O abundance model. 
For the case of decreased abundances, we do not see any noticeable differences from the case of decreased C and O abundances. 
This result, and that of the case in which only the Fe abundance is varied indicate that the spectrum of the T dwarf is not affected by Fe dust.

\section{Fitting the {\COt} 4.2~$\mu$m Absorption Band with Models of Different Elemental Abundances}
\begin{figure}
\begin{center}
\epsscale{.5}
\plotone{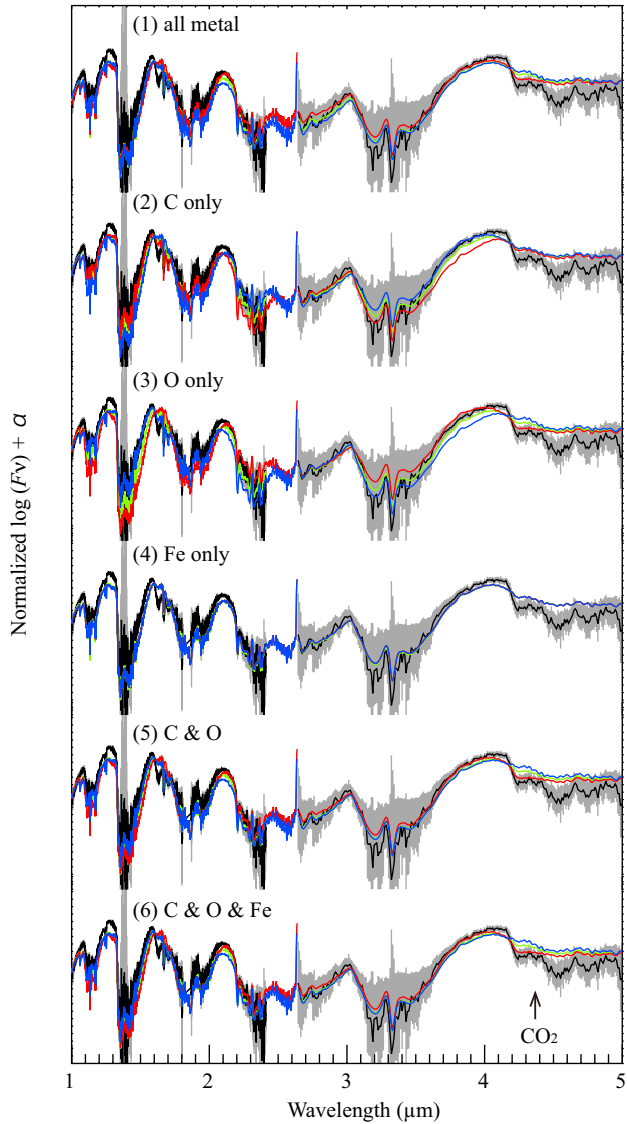}
\end{center}
\caption{Observed and model spectra of 2MASS J0559--1404(T4.5). Grey lines indicate errors. 
The model spectra in which elemental abundance are varied are drawn in colors.
Green: the model spectra of the solar abundance, red: increased, and blue: decreased elemental abundances. 
We construct models where the following parameters are varied: 
(1) all metal abundances, (2) only C abundance, (3) only O abundance, (4) only Fe abundance, (5) C and O abundances, and (6) C and O and Fe abundances. 
The spectral shapes, especially of the {\COt} 4.2~$\mu$m band, in the model spectra of increasing $``$C \& O$"$, and $``$C \& O \& Fe$"$ fits the observation quite well without changing other features.
} 
\label{com}
\end{figure}

In order to see the effects of elemental abundances on the infrared spectra of brown dwarfs more clearly, 
we calculate models of various elemental abundances for the best-fit model-parameter-set  
determined by \citet{Sorahana_2012} for {\AKARI} + IRTF/SpeX\footnote{We obtained the SpeX data  from the SpeX Prism Spectral
Libraries built by Dr. Adam Burgasser and Dr. Sandy Leggett (http://pono.ucsd.edu/$\sim$adam/browndwarfs/spexprism/html/all.html) and the IRTF Spectral Library maintained by Dr. Michael Cushing (http://irtfweb.ifa.hawaii.edu/$\sim$spex/IRTF\_Spectral\_Library/).} and UKIRT/CGS4\footnote{We obtained the spectral data of SDSS J1446+0024 from Dr. Dagny Looper (2010, private communication).} spectra.
Then we compare the result with the observed data and the solar abundance models.
In this section we mainly discuss how the {\COt} band behaves under different elemental abundances. 
The six cases of modified elemental abundances examined in Section~\ref{s2}  
are considered. 
We show a comparison of the observed spectrum and these six model spectra in Figure~\ref{com}. 
We choose 2MASS~J0559--1404 (T4.5) as an example, because the object shows the {\COt} absorption band clearly in its high signal-to-noise ratio spectrum. 
The {\COt} 4.2~$\mu$m absorption band in the observed spectrum of this object is deeper than that of the best-fit model of the solar abundance. 
The {\COt} absorption band in the model spectra of varied only C abundance  and only Fe abundance show minor changes in the {\COt} band region.  
For the case of only increasing O abundance, we see that the {\COt} 4.2~$\mu$m feature in the model spectra is significantly deepened, 
but as a side effect other features in the model spectra, for example {\HtO} around 1.4, 1.8, and 2.7~$\mu$m and {\CHf} at 3.0--4.0~$\mu$m, deviate from the observation. 
The {\COt} 4.2~$\mu$m band in the model spectra of increased $``$C and O$"$ abundances fits the observation reasonably well without changing other features. 
We find that the O abundance plays the largest role in the photosphere of brown dwarfs, 
and the C abundance controls the extra chemical effects caused by the O abundance. 
This result indicates that the ratio of C to O in this brown dwarf atmosphere is similar to that of the sun.
The  model in which $``$C and O and Fe$"$ are varied behaves similarly to the cases in which both $``$C and O$"$ are varied,  
in the case of a T4.5 dwarf.  
The behavior of spectral features for the case of increased all metal abundances is also similar to the case in which $``$C and O (and Fe)$"$ abundances are varied.
This indicates that the most effective elements to reproduce the near-infrared spectral features of brown dwarfs are C and O, 
even though other elements such as Fe also contributes to suppress the variation occurred by C and O abundances as discussed in Section 2.5 and 2.6.
To constrain the other elemental abundances, 
we should investigate other band and line features, such as CO, {\CHf}, K~I, and FeH, with high resolution and high sensitivity spectroscopy in  future.

\begin{figure*}
\begin{center}
\epsscale{.3}
\plotone{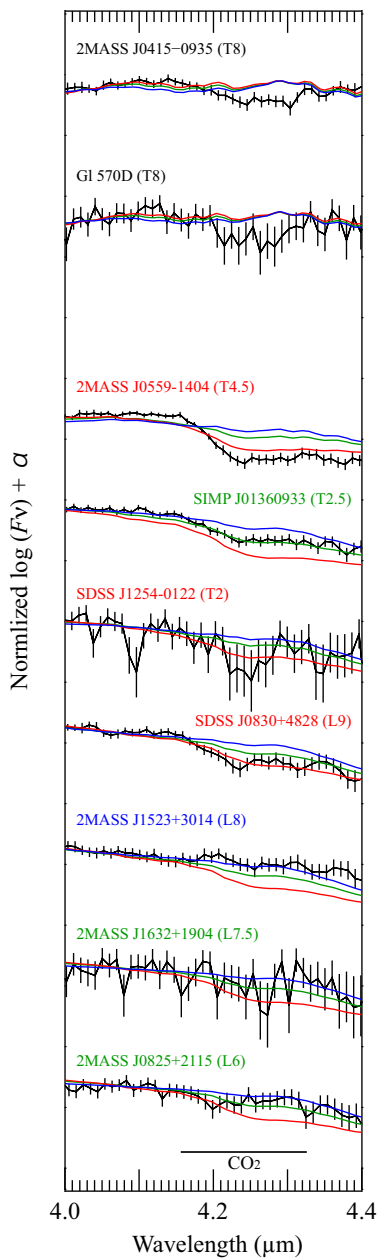}
\end{center}
\caption{Normalized observed and model spectra of mid-L to late-T dwarfs  in 4.0--4.4~$\mu$m.. Black lines are observed spectra, and green lines are their best fit model spectra with the solar abundance. Red and blue lines are models of increased and decreased C and O elemental abundances, respectively. The {\COt} 4.2~$\mu$m absorption feature in the spectra of three objects with names written in red are better  explained by the model with increased C and O elemental abundances, and the object whose name is written in blue is well reproduced by the decreased abundances model. 
The spectra of the other three sources labelled in green are best explained by the solar abundance model. } 
\label{ocom}
\end{figure*}

Next, we calculate models in which $``$C and O$"$ abundances are varied for other middle- to late-L and T dwarfs  
and check the behavior of the {\COt} 4.2~$\mu$m fundamental absorption band.  
Since the effect of Fe on spectral features is relatively minor compared to that of C and O, we only consider the models in which $``$C and O$"$  abundances are varied following \citet{Tsuji_2011}.
The results are shown in Figure~\ref{ocom}. 
We find that the {\COt} 4.2~$\mu$m absorption band does not change at all by changing elemental abundance for late-T dwarfs ({\Teff} $\sim$ 700~K).  
In the models of late-T dwarfs, the {\COt} abundance is already so low and small changes of the abundance will not change the spectrum. 
The presence of the {\COt} band in these sources should be explained by another mechanism.
We find that the {\COt} band in the spectra of the models with increased $``$C and O$"$ abundances fits the spectra of 2MASS~J0559--1404 and 2MASS~J0830+4828  better than the solar abundance models, 
as concluded by \citet{Tsuji_2011}. 
The model with increased $``$C and O$"$ abundances also fits the spectra of SDSS J1254--0122 better than the solar abundance model, although the error of the observed spectrum is large. 
On the other hand, the model of decreased $``$C and O$"$ gives better fits for 2MASS J1523+3014 than the original model. 
This result of 2MASS J1523+3014 being metal-poor is consistent with the statement by \citet{Kirkpatrick_2001}. 
This object is a companion of well known G-type star.
Thus, they suggested that metallicity of 2MASS J1523+3014 (GJ~584C called in their paper) is --0.2~dex assigned from that of the primary star.
We for the first time confirm the elemental abundances of this brown dwarf directly and also confirm that the C and O abundances of the companion is almost same with those of the primary star.
We also confirm that the model of decreased all metal abundances does not give the better fit for the {\COt} band because the variation is suppressed by other elements except for C and O.
Our results imply that 2MASS J0559--1404, SDSS J0830+4828 and SDSS J1254--0122 are C and O-rich relative to the Sun, and 2MASS J1523+3014 is a C and O poor object. 

\section{Summary}
We attempt to improve brown dwarf atmosphere models by varying elemental
abundances. 
We calculate the model atmospheres of typical L and T dwarfs varied with (1) all metal abundances (elements except for H and He); (2) C abundance
only; (3) O abundance only; (4) Fe abundance only; (5) C and O abundances; and (6) C and O and Fe abundances. 
We investigate the variation of CO, {\COt}, {\CHf} and {\HtO} abundances and temperature in the photospheres. 
We also examine flux levels at several wavelengths in the spectra, namely the $J$, $H$, and $K$ bands, the centers of CO bands at 4.6~$\mu$m,  the {\COt} band at 4.2~$\mu$m, the {\CHf} band at 3.3~$\mu$m, and the {\HtO} bands at 1.4, 1.8 and 2.7~$\mu$m. 

We now summarize the variation of molecular abundances. 
In the case of only increasing the  C  abundance, the CO and {\CHf}  abundances increase.
Since more O atoms are captured in CO,
the {\HtO} and {\COt} abundances decrease. 
However, the variations of {\HtO} and {\COt} abundances in T dwarf photospheres are smaller than those in L dwarfs.
This is because of a small variation of CO abundance in T dwarf photosphere.
In all cases of increasing O abundance, the abundances of {\HtO} and {\COt} in the photosphere increase. 
All decreasing cases behaves opposite to the increasing case.

Temperature generally correlates tightly with {\HtO} and {\COt} abundances. 
When the {\HtO} and/or {\COt} abundances increase in the optically thick region of the photosphere, they play a role of holding the energy from the inner photosphere  and
the temperature in the region rises (greenhouse effect).  
On the other hand, when the {\HtO} and/or {\COt} abundances in the optically thin region increase, 
the temperature in the region decreases through radiative cooling. 
The temperature around dust layers depends on the amount of dust, especially of iron grains.  
Fe has the largest extinction effect relative to other dust components, Al$_2$O$_3$ and MgSiO$_3$, included in the UCM. 
When Fe increases, the temperature around the layers increases because of the greenhouse effect. 
The {\CHf} abundance tends to be affected by the variations of the photosphere temperature in addition to the C abundance. 

In general, the flux level of each molecular band reflects its abundance and temperature profile along the line of sight. 
The three T dwarf models where the following parameters are varied: (1) all metal abundances, (2) C and O abundances, and (3) C and O and Fe abundances, 
show spectra similar to each other, 
and only the {\COt} absorption band changes noticeably.
The flux levels of the $J$, $H$, and $K$ bands reflect the temperature of the inner photosphere where Rossland and Planck mean optical depth$\tau_{R} \sim1$ and {\HtO} abundance.  
The $J$ and $H$ bands are also sensitive to the dust abundance.

\citet{Sorahana_2012} have found that the observed {\COt} absorption bands in some objects are stronger or
weaker than the prediction by the solar abundance models. 
We construct a set of model atmospheres with various elemental abundances for the same model parameters determined by \citet{Sorahana_2012}, and compare the model spectra with the observed spectra.
First, we compare the observed spectrum of 2MASS~J0559--1404(T4.5) with six model spectra with various elemental abundances.
As a result, the {\COt} band in the model spectrum with increased $``$C and O (and Fe)$"$ abundances and increased all metal abundances fit observed spectra better than that of the solar abundance models and any other abundance pattern models.
This indicates that C and O are the most effective elements to reproduce the near-infrared spectral features of brown dwarfs.
Next, we attempt to investigate whether the {\COt} absorption band in the spectrum of late-L to T dwarfs are explained by the model in which only C and O abundances are varied.
We find that the excess of {\COt} abundance in the observed spectra of three objects can be reproduced by the increased abundance model,
and there is one object for which the model in which C and O abundances are decreased better explains the band in the observed spectrum. 
This indicates that there are C and O-poor brown dwarfs. 
We also confirm that the model of decreased all metal abundances does not fit better, at least, for 2MASS~J1523+3014 because the variation of {\COt} band strength is suppressed by other elements except for C and O.
This may suggests that C and O abundances vary independently of other elements.
The result of poor C and O abundances in 2MASS~J1523+3014 is consistent with the statement of \citet{Kirkpatrick_2001}.
The remaining three objects are best fit by the solar abundance model.
These results indicate that both C and O abundances should increase and decrease simultaneously. 

Recently, \citet{Madhusudhan_2011} reported an anomaly in the {\HtO} and {\CHf} abundances compared to the solar abundance chemical equilibrium model prediction for the atmosphere of the hot-Jupiter WASP-12b. 
They suggested that the abundance of these molecules can be explained if the carbon-to-oxygen ratio C/O in this planet's atmosphere is much greater than the solar value (C/O $=0.54$), C/O $\ge 1$ at $3\sigma$ significance. 
From our result, however, we suggest that the ratio of C to O in our {\AKARI} brown dwarf sample should be closer to the solar value. 
This difference between brown dwarfs and exoplanet is potentially caused by their formation. 
While, almost all isolated brown dwarfs are born in the interstellar medium like stars,
planets are born in protopranetaly disks. 
There are several mechanisms for planet formation, e.g. core accretion (CA) and gravitational instability (GI).
In CA model, giant gas planets have a lot of {\HtO} in their core, because the core first forms from a {\HtO} rich disk.
After the core forms, they capture a lot of gas from surrounding disk with lack of {\HtO}, i.e., O.
Thus the giant gas planets formed by the CA may have a O-poor atmosphere.
On the other hand, the giant gas planets in GI model form directly in mixed gas and dust, 
thus it is considered that their atmospheres have no lack of O.
WASP-12b is possibly formed by the CA.
If most of planets are formed by CA, C/O ratio will be an index to distinguish between brown dwarf and planet.

The range of elemental abundances (from --0.2 to +0.2 dex) is within the range of metallicity variation among solar neighborhood stars (i.e. [X/H] distributed between $-$1.0 and 1.0~dex). 
Our analysis also reveals a possibility that a remarkable fraction of brown dwarfs seem to be C and O-rich. 
On the other hand, none of the revised models can explain the {\COt} absorption band strength in the latest T dwarfs yet. 
To understand the deviation of the {\COt} abundance in these T dwarf photospheres, 
we need to consider other mechanisms.

We thank the anonymous referee for critical reading of our article and for invaluable suggestions.
We acknowledge Dr. Adam Burgasser, Dr. Sandy Leggett and Dr. Michael. Cushing for providing us observed near-infrared spectral data.
This research is based on observations with {\AKARI}, a JAXA project with the participation of ESA. 
We thank  Prof. Takashi Tsuji for his kind permission to access the UCM and for helpful suggestions. 
We also thank Takafumi. Kamizuka for useful discussions throughout this research.
We are grateful to Dr. Jennifer. Stone for her careful checking of the manuscript and many suggestions to improve the text.
This work was supported by Grants-in-Aid for Scientific Research
from the Japan Society for the Promotion of Science (JSPS).
This work is supported by JSPS/KAKENHI(c) No. 22540260 (PI: I. Yamamura) and in part by Grants-in-Aid for Scientific Research from the MEXT of Japan, 22864006 (PI: T.K.Suzuki). 




\bibliography{sorahana}

\begin{thebibliography}{25}
\expandafter\ifx\csname natexlab\endcsname\relax\def\natexlab#1{#1}\fi

\bibitem[{{Allende Prieto} {et~al.}(2002){Allende Prieto}, {Lambert}, \&
  {Asplund}}]{Allende_2002}
{Allende Prieto}, C., {Lambert}, D.~L., \& {Asplund}, M. 2002, ApJ, 573, L137

\bibitem[{{Anders} \& {Grevesse}(1989)}]{Anders_1989}
{Anders}, E., \& {Grevesse}, N. 1989, \gca, 53, 197

\bibitem[{{Barman} {et~al.}(2011){Barman}, {Macintosh}, {Konopacky}, \&
  {Marois}}]{Barman_2011}
{Barman}, T.~S., {Macintosh}, B., {Konopacky}, Q.~M., \& {Marois}, C. 2011,
  \apj, 733, 65

\bibitem[{{Bodaghee} {et~al.}(2003){Bodaghee}, {Santos}, {Israelian}, \&
  {Mayor}}]{Bodaghee_2003}
{Bodaghee}, A., {Santos}, N.~C., {Israelian}, G., \& {Mayor}, M. 2003, A\&A,
  404, 715

\bibitem[{Chackerian \& Tipping(1983)}]{Chackerian_1983}
Chackerian, C.~J., \& Tipping, R.~H. 1983, J. Mol. Spectrosc., 99, 431

\bibitem[{{Fortney}(2012)}]{Fortney_2012}
{Fortney}, J.~J. 2012, \apjl, 747, L27

\bibitem[{Freedman {et~al.}(2008)Freedman, Marley, \& Lodders}]{Freedman_2008}
Freedman, R.~S., Marley, M.~S., \& Lodders, K. 2008, ApJS, 174, 504

\bibitem[{{Grevesse} {et~al.}(1991){Grevesse}, {Lambert}, {Sauval}, {van
  Dishoeck}, {Farmer}, \& {Norton}}]{Grevesse_1991}
{Grevesse}, N., {Lambert}, D.~L., {Sauval}, A.~J., {et~al.} 1991, \aap, 242,
  488

\bibitem[{{Griffith} \& {Yelle}(1999)}]{Griffith_1999}
{Griffith}, C.~A., \& {Yelle}, R.~V. 1999, ApJ, 519, L85

\bibitem[{Guelachivili {et~al.}(1983)Guelachivili, De~Villeneuve, Farrenq,
  Urban, \& Verges}]{Guelachvili_1983}
Guelachivili, G., De~Villeneuve, D., Farrenq, R., Urban, W., \& Verges, J.
  1983, J. Mol. Spectrosc., 98, 64

\bibitem[{{Kirkpatrick} {et~al.}(2001){Kirkpatrick}, {Dahn}, {Monet}, {Reid},
  {Gizis}, {Liebert}, \& {Burgasser}}]{Kirkpatrick_2001}
{Kirkpatrick}, J.~D., {Dahn}, C.~C., {Monet}, D.~G., {et~al.} 2001, \aj, 121,
  3235

\bibitem[{{Konopacky} {et~al.}(2013){Konopacky}, {Barman}, {Macintosh}, \&
  {Marois}}]{Konopacky_2013}
{Konopacky}, Q.~M., {Barman}, T.~S., {Macintosh}, B.~A., \& {Marois}, C. 2013,
  Science, 339, 1398

\bibitem[{{Madhusudhan} {et~al.}(2011){Madhusudhan}, {Harrington}, {Stevenson},
  {Nymeyer}, {Campo}, {Wheatley}, {Deming}, {Blecic}, {Hardy}, {Lust},
  {Anderson}, {Collier-Cameron}, {Britt}, {Bowman}, {Hebb}, {Hellier},
  {Maxted}, {Pollacco}, \& {West}}]{Madhusudhan_2011}
{Madhusudhan}, N., {Harrington}, J., {Stevenson}, K.~B., {et~al.} 2011, \nat,
  469, 64

\bibitem[{{Moses} {et~al.}(2013){Moses}, {Madhusudhan}, {Visscher}, \&
  {Freedman}}]{Moses_2013}
{Moses}, J.~I., {Madhusudhan}, N., {Visscher}, C., \& {Freedman}, R.~S. 2013,
  \apj, 763, 25

\bibitem[{{Murakami} {et~al.}(2007){Murakami}, {Baba}, {Barthel}, {Clements},
  {Cohen}, {et~al.}}]{Murakami_2007}
{Murakami}, H., {Baba}, H., {Barthel}, P., {et~al.} 2007, PASJ, 59, 369

\bibitem[{Partridge \& Schwenke(1997)}]{Partridge_1997}
Partridge, H., \& Schwenke, D.~W. 1997, J. Chem. Phys., 106, 4618

\bibitem[{{Rothman}(1997)}]{Rothman_1997}
{Rothman}, L.~S. 1997, {High-temperature Molecular Spectroscopic Database
  (CD-ROM)} (Andover: ONTAR Co.)

\bibitem[{{Saumon} {et~al.}(2000){Saumon}, {Geballe}, {Leggett}, {Marley},
  {Freedman}, {et~al.}}]{Saumon_2000}
{Saumon}, D., {Geballe}, T.~R., {Leggett}, S.~K., {et~al.} 2000, ApJ, 541, 374

\bibitem[{{Sorahana} \& {Yamamura}(2012)}]{Sorahana_2012}
{Sorahana}, S., \& {Yamamura}, I. 2012, ApJ, in press.

\bibitem[{{Swain} {et~al.}(2009){Swain}, {Vasisht}, {Tinetti}, {Bouwman},
  {Chen}, {Yung}, {Deming}, \& {Deroo}}]{Swain_2009}
{Swain}, M.~R., {Vasisht}, G., {Tinetti}, G., {et~al.} 2009, \apjl, 690, L114

\bibitem[{{Tsuji}(2002)}]{Tsuji_2002}
{Tsuji}, T. 2002, ApJ, 575, 264

\bibitem[{{Tsuji}(2005)}]{Tsuji_2005}
---. 2005, ApJ, 621, 1033

\bibitem[{{Tsuji} {et~al.}(2011){Tsuji}, {Yamamura}, \&
  {Sorahana}}]{Tsuji_2011}
{Tsuji}, T., {Yamamura}, I., \& {Sorahana}, S. 2011, ApJ, 734, 73

\bibitem[{{Wenger} \& {Champion}(1998)}]{Wenger_1998}
{Wenger}, C., \& {Champion}, J.~P. 1998, \jqsrt, 59, 471

\bibitem[{{Yamamura} {et~al.}(2010){Yamamura}, {Tsuji}, \&
  {Tanab{\'e}}}]{Yamamura_2010}
{Yamamura}, I., {Tsuji}, T., \& {Tanab{\'e}}, T. 2010, ApJ, 722, 682

\end{thebibliography}

\end{document}